\documentclass[a4paper,11pt]{article}

\usepackage{pos}
\usepackage{lipsum} 
\usepackage[normalem]{ulem}
\usepackage{soul}
\usepackage{wrapfig}

\title{The GRAND@Auger Prototype for the Giant Radio Array for Neutrino Detection}

\ShortTitle{GRAND@Auger}

\author[a]{Beatriz de Errico}
\author*[a]{Jo\~ao R. T. de Mello Neto}
\author[b,c]{Charles Timmermans}
\onbehalf{for the GRAND Collaboration{\normalsize \\ \normalfont(a complete list of authors can be found at the end of the proceedings)}\\}

\affiliation[a]{Instituto de Física, Universidade Federal do Rio de Janeiro - UFRJ ,\\
  Ilha do Fund\~ao , Rio de Janeiro , Brazil }
\affiliation[b]{Institute for Mathematics, Astrophysics and Particle Physics, Radboud Universiteit, Nijmegen, the Netherlands}
\affiliation[c]{Nikhef, National Institute for Subatomic Physics, Amsterdam, the Netherlands}

\emailAdd{beatrizspe@pos.if.ufrj.br}
\emailAdd{jtmn@if.ufrj.br}
\emailAdd{c.timmermans@science.ru.nl}

\abstract{

As of early 2025, the GRAND Collaboration operates three prototype arrays: GRANDProto300 in China, GRAND@Nan\c cay in France, and GRAND@Auger in Argentina. The GRAND@Auger prototype was established through an agreement  between the GRAND and Pierre Auger Collaborations, repurposing ten Auger Engineering Radio Array (AERA) stations as GRAND detection units. This setup provides a unique opportunity for coincident air-shower detection, enabling direct event-by-event comparison between GRAND@Auger and the well-established Pierre Auger detectors. Such comparisons allow for a detailed assessment of the detection principle and reconstruction capabilities of GRAND.  In this contribution, we present an overview of the commissioning and preliminary results of GRAND@Auger, including the measurement of the Galactic background noise and the detection of the first  self-triggered candidate event in coincidence with the Pierre Auger Observatory. Both results underscore the role of GRAND@Auger in advancing the GRAND project and refining the techniques required for large-scale radio detection of ultra-high-energy particles.

\vspace{4mm}

}

\ConferenceLogo{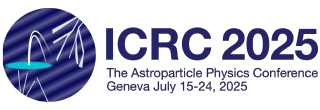}

\FullConference{39th International Cosmic Ray Conference (ICRC2025)\\  15--24 July, 2025\\ Geneva, Switzerland}

\begin{document}

\maketitle

\section{Introduction}

The Giant Radio Array for Neutrino Detection (GRAND)~\cite{GRAND:2018iaj,Olivier:ICRC25} is a proposed large-scale observatory designed to identify and study the sources of ultra-high-energy cosmic rays (UHECRs). A central goal of GRAND is the detection and characterization of ultra-high-energy (UHE) neutrinos. The proposed radio-antenna array will operate by detecting radio signals produced when UHE cosmic rays, gamma rays, and neutrinos induce extensive air showers (EAS) in Earth's atmosphere. In addition, its design enables a broad range of studies in fundamental particle physics, cosmology, and radio astronomy. GRAND is expected to make significant contributions to the detection of neutrino emissions from transient astrophysical sources~\cite{Guepin:2022qpl}. 

The GRAND Collaboration has deployed three prototype arrays: GRAND@\allowbreak Nan\c cay in France~\cite{Correa:ICRC2025}, GRANDProto300 in China~\cite{Ma:ICRC2025}, and GRAND@Auger (G@A) in Argentina.  These pave the way for self-triggered radio-antenna arrays capable of detecting very inclined air showers through their radio emission, a GRAND's detection concept requirement. In this paper, we focus on the G@A prototype.

\begin{figure}[htbp]
    \centering
    \begin{minipage}{0.485\textwidth}
        \centering
        \includegraphics[width=\linewidth]{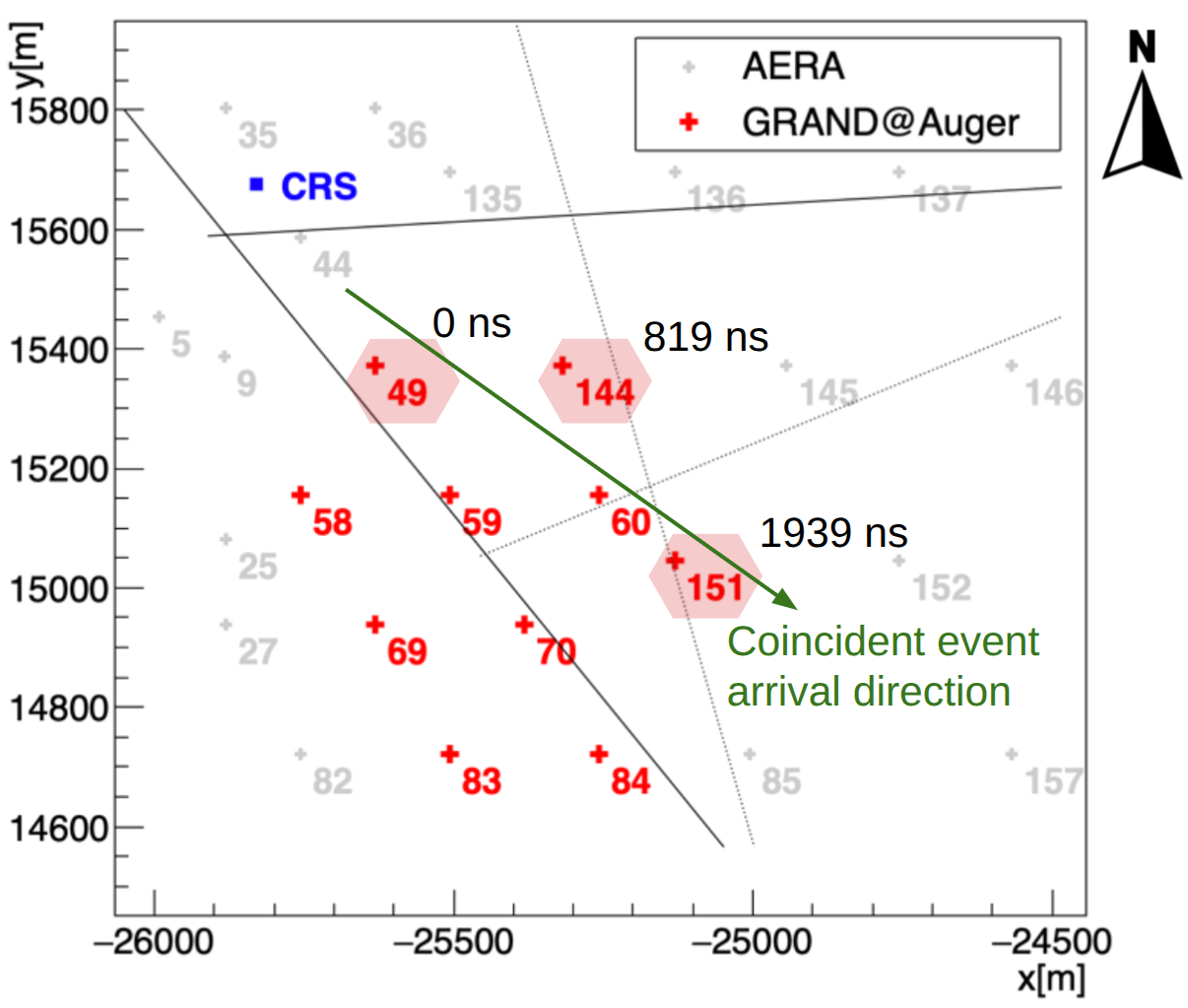}
        \caption{The layout of the G@A prototype is shown in red, with nearby AERA stations indicated in grey. The lines represent local roads. The Central Radio Station, which houses the central data acquisition computer, is highlighted in  blue. The numbers in black are the timestamps  in nanoseconds, relative to detection unit 49, for the coincident event that will be presented in section \ref{sec:event}.}
        \label{fig:array}
    \end{minipage}
    \hfill
    \begin{minipage}{0.485\textwidth}
        \centering
        \includegraphics[height =0.8\linewidth ]{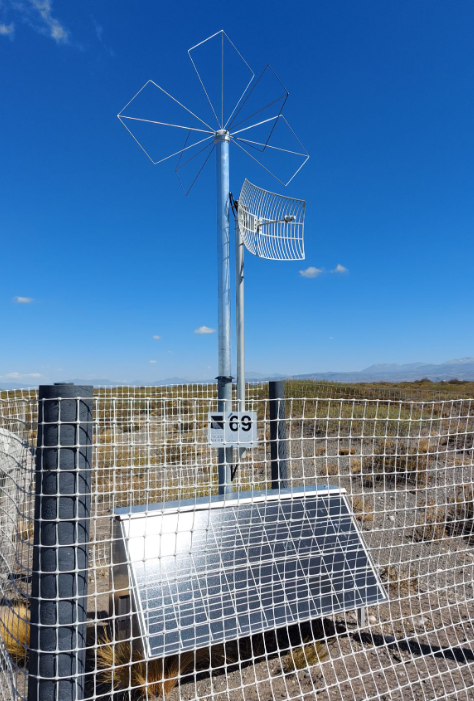}
        \caption{Image of one of the converted antennas at the G@A site. The 3-meter pole supports the antenna and the nut containing the LNA. Midway up the pole, the mesh communication antenna is installed. 
         The space below the solar panel houses the battery, charge controller and front-end electronics.
        The entire setup is enclosed by a plastic fence.}
        \label{fig:antenna}
    \end{minipage}
\end{figure}

\section{The GRAND@Auger prototype} 

The G@A prototype represents a  joint effort between the GRAND and Pierre Auger  Collaborations  to validate the GRAND detection concept for ultra-high-energy air showers using self-triggered radio antennas. The G@A prototype array is located at the Pierre Auger Observatory in Malarg\"ue, Argentina, as shown in Figure \ref{fig:array}. Ten butterfly antenna stations from the Auger Engineering Radio Array (AERA)~\cite{deJong2017} (located in the Auger's Surface detector 750-m array)  were repurposed into GRAND detection units.  An array layout consisting of two superimposed hexagons with a spacing of approximately 250~m was selected. The resulting array covers an area of roughly 0.5~km$^2$. Space at the Pierre Auger Observatory's Central Radio Station (CRS) was allocated to house the G@A Central DAQ computer. The dense layout configuration was chosen to  enhance
the probability of detecting lower-energy and less inclined events.

 A photo of a GRAND@Auger detection unit is shown in Figure~\ref{fig:antenna}. Each GRAND detection unit consists of a Horizon Antenna mounted on a 3-meter pole, measuring radio signals in three polarizations: North-South (NS), East-West (EW), and Vertical (V). The mechanical structure retains some of the AERA infrastructure, including the solar panels and communication antenna, while integrating new GRAND-specific components. The Horizon Antenna is supported by a nut at the top of the pole, which also houses the low noise amplifier (LNA). Midway along the pole, the GPS and wireless communication modules are installed. At the base, a weather-sealed aluminum box contains the front-end electronics  board, a battery, and a charge controller, all enclosed within Faraday cages to mitigate self-induced electromagnetic noise. 

\begin{figure}[t]
 \centering
 \includegraphics[width=0.7\textwidth]{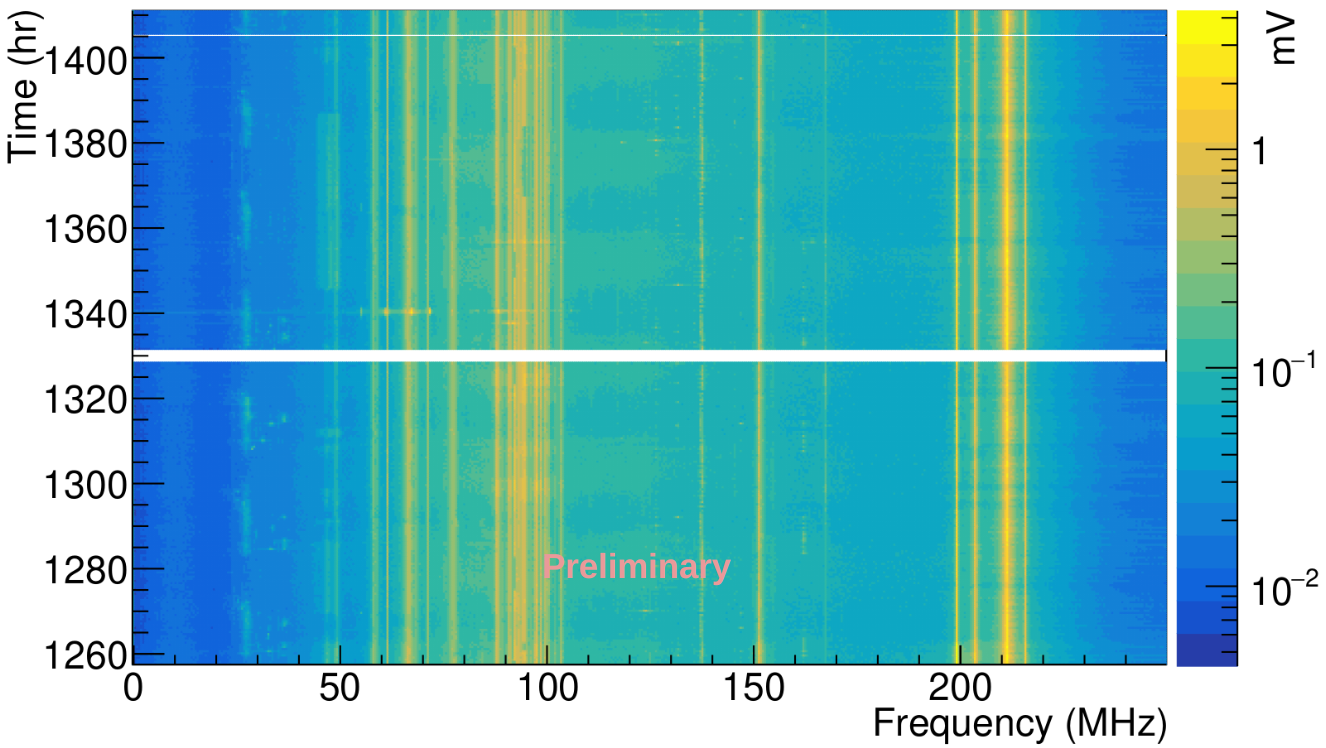}
 \caption{ Periodogram of six days of data from detection unit 59. Constant radio background signals appear as vertical yellow lines, while noise-free regions are also visible. The white horizontal lines indicate hibernation periods, during which the detection unit temporarily ceased data acquisition and later resumed automatically when conditions permitted.  }
   \label{fig:periodogram}
 \end{figure}

The electronic architecture includes a 500 MSPS ADC chip for digitization, combined with a Xilinx SoC processor -- containing the FPGA and the CPU chips-- that performs flexible filtering and triggering. The system implements a 30--200~MHz bandpass filter and applies periodic or threshold-based triggers. Data acquisition is handled locally by the on-board processor, which  communicates with the central data acquisition system and sends the requested event information. These events are transmitted via a wireless \texttrademark{Ubiquiti} Bullet-Rocket system to the central data acquisition computer at the Central Radio Station. The central system  constructs the triggered events and writes them into disk, which are then transferred daily to the CCIN2P3 data center in Lyon for analysis.

 GRAND Collaboration teams, aided by Pierre Auger Observatory local staff, deployed the G@A prototype array in two campaigns in March and August 2023, then followed by commissioning efforts. By March 2024, the ten G@A detection units were operational. Data taking was stable, and hibernation settings triggered by either high temperature or low battery levels were functional. An example periodogram from one detection unit, containing nearly 150 hours of continuously recorded periodically triggered data, is shown in Figure~\ref{fig:periodogram}.
Further details on the GRAND@Auger analysis and first results can be found in~\cite{deErrico:2024}.

\section{Galactic Calibration Analysis}
\label{sec:gal}

 The GRAND Collaboration makes on-going modifications and updates to the hardware and software of its prototype arrays detection units. The Galactic background analysis for the G@A prototype is based on data collected from March to June 2024, a period of stable hardware configuration. The dataset comprises measurements from the ten detection units, but careful selection and filtering were necessary to ensure data quality.

\begin{figure}[t]
 \centering
 \includegraphics[width=.95\textwidth, trim={.6cm 0 2cm 1.2cm}, clip]{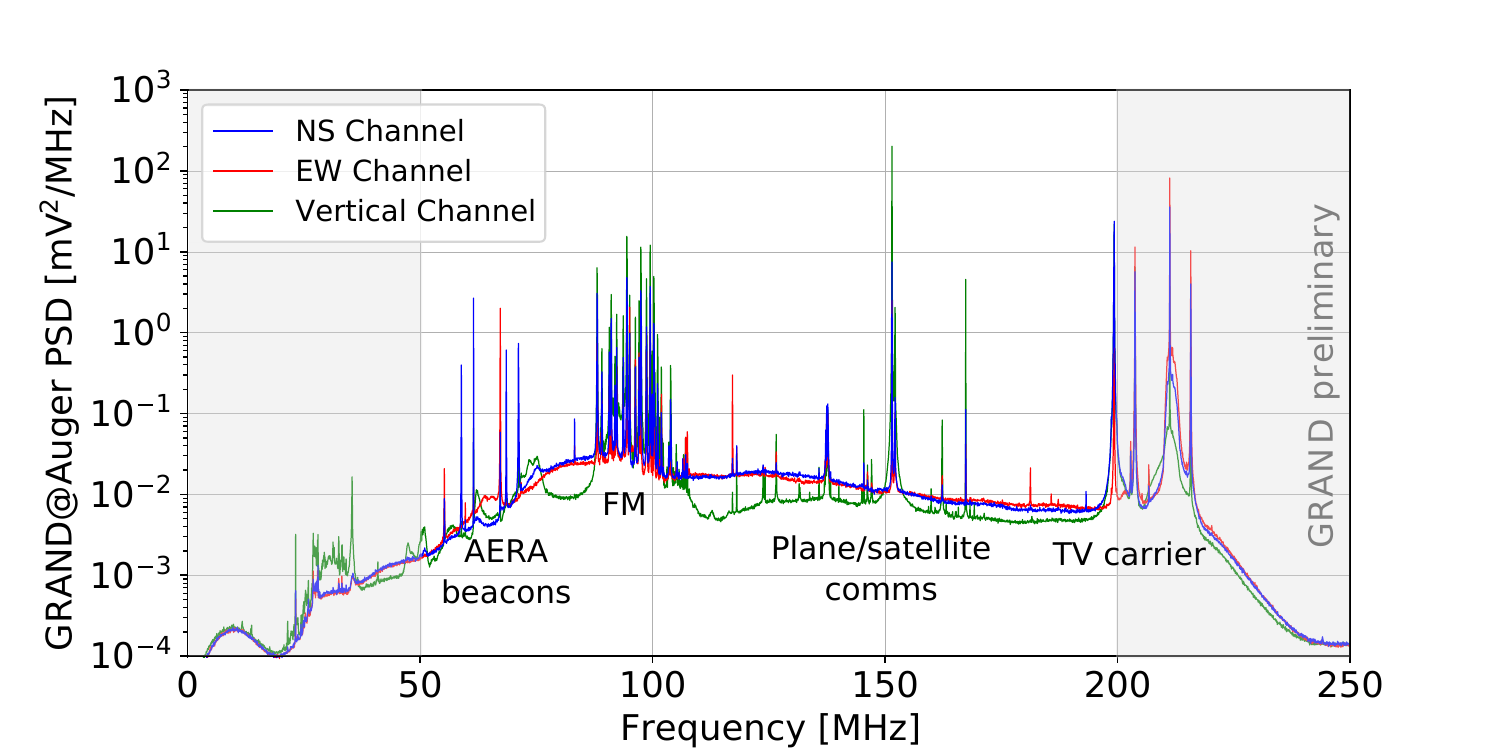}
 \caption{ Average power spectral density evaluated at three polarizations over 8 hours of data from DU~83 in March~2024. The data used are from a special run of 16~$\mu$s-long time series with a frequency resolution of approximately 62.5~kHz. Some of the known sources of the power spectrum density peaks are indicated below the PSD curves.}  \label{fig:psd}
\end{figure}

The data selection addressed the following issues:

\begin{itemize}
    
    \item \textbf{Local Noise and Station-Specific Anomalies:}  
     In some stations the electronics sporadically shows issues on gain and noise. After removal of these periods, five  detection units (58, 59, 69, 144, 151) still exhibited anomalous behavior in several channels. These  were excluded, leaving five  detection units (49, 60, 70, 83, 84) for the Galactic analysis.
    
    \item \textbf{Single-Frequency Sources:} Narrowband radio frequency interference (e.g., video carriers, aviation communications, and  satellite signals, as shown in Figure \ref{fig:psd}) was identified and removed through targeted frequency cuts, ensuring a clean broadband dataset. The filtering is done  in two stages: online at the FPGA level using a programmable digital notch filter  and offline at the event analysis level.
\end{itemize}

\begin{wrapfigure}{R}{0.5\textwidth}
    \centering
    \includegraphics[width=\linewidth, trim={9.5cm 8.9cm 2.7cm 1.9cm}, clip]{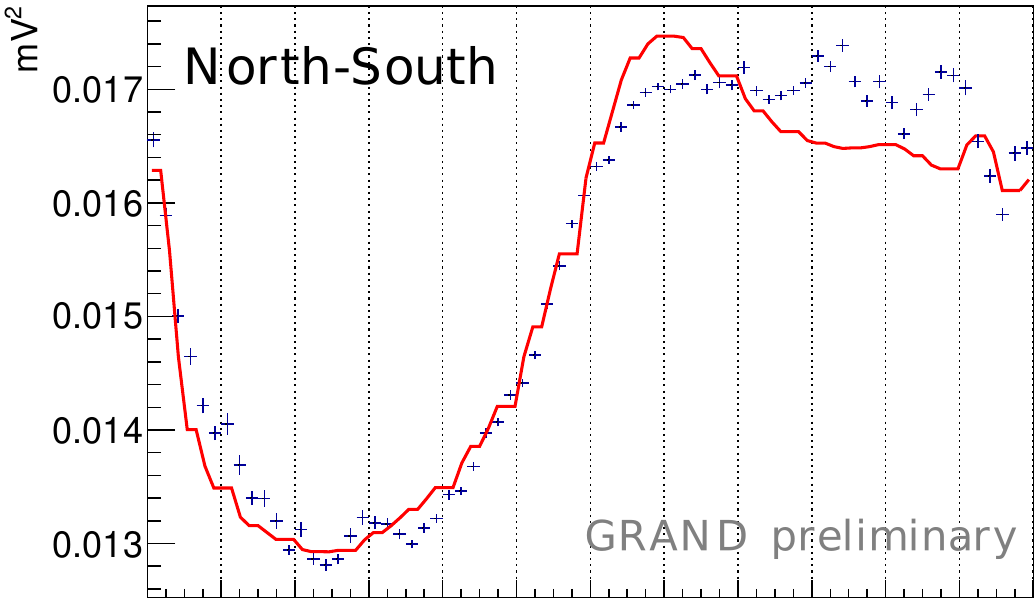}
    \includegraphics[width=\linewidth, trim={9.5cm 7.7cm 2.7cm 1.9cm}, clip]{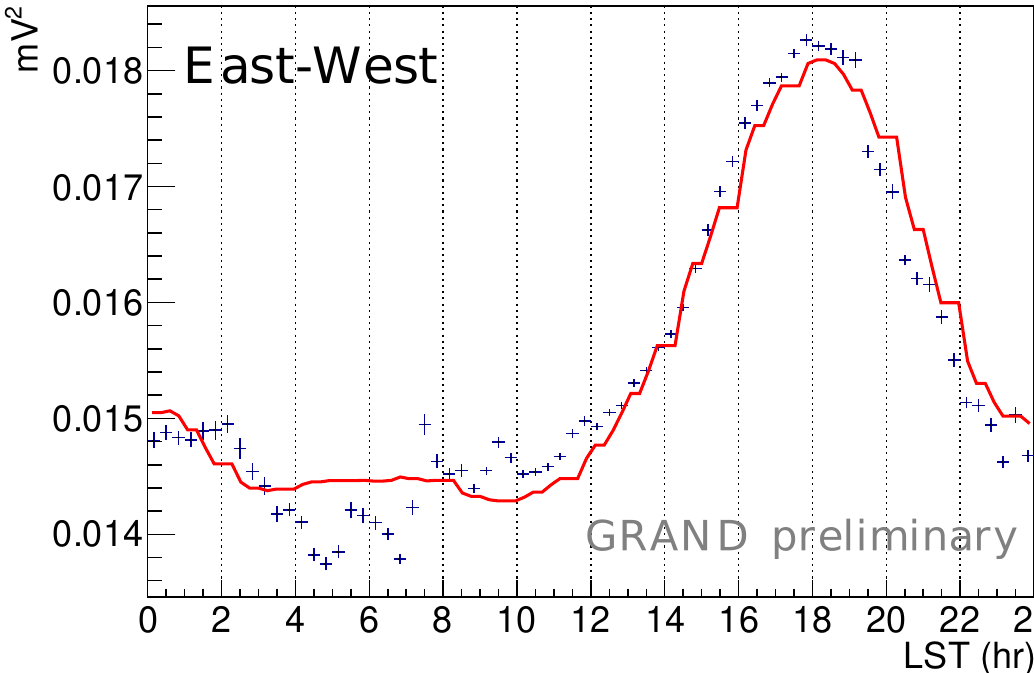}
   \caption{Data from  detection unit 83 in blue, binned in 20-minute LST intervals, each line corresponds to an antenna polarization. The error bars indicate the uncertainty on the mean. The red line represents the galactic simulation fitted to the data, with an additional offset and amplification. Only the power above 100 MHz is used.
   }
    \label{fig:bgal}
\end{wrapfigure}

After applying all cuts, approximately 50 days of data remained for the galactic  background analysis. The resulting time-averaged power spectra showed periodic features, but the limited data duration prevented a clear separation between signals following the Local Sidereal Time (LST) and those following the 24-hour GMT cycle. Longer datasets are necessary for a definitive distinction.

To interpret the measured signals, comparisons were made with simulations modeling the expected Galactic signal. The Galactic background predictions used in this analysis were produced using the \texttt{LFmap} program~\cite{Polisensky:2007}, which generates low-frequency sky maps for radio astronomy applications.  The simulated emissions were folded with the GRAND radio frequency chain (RF chain) and were added with random Gaussian noise -- implemented in the {\sc GRANDlib} software \cite{GRANDlib2025}. This procedure emulates the Galactic signals as if they were detected by a GRAND detection unit. A linear model was fitted to the measured power:
\[
\text{Signal(LST)} = \text{Offset} + \text{Gain} \times \text{Simulation(LST)},
\]

\noindent with the gain and offset optimized for each frequency range and  detection unit. The analysis focused on three frequency bands: 30--200 MHz, 30--100 MHz, and 100--200 MHz. The comparison revealed:

\begin{itemize}
    \item Better agreement at higher frequencies (100--200 MHz), where the gain calibration is understood to approximately 10\% precision for horizontal polarizations and up to 50\% for vertical channels.
    \item Weaker correlation at lower frequencies (30--100 MHz),  which was also present in the full frequency range analysis (30--200 MHz).  This indicates limitations in the current understanding of the system response.
\end{itemize}

The resulting data and simulation are shown in Figure~\ref{fig:bgal},  in which there is reasonable agreement between both. The galactic center should be turned to Malargüe at around 16-18 hours in local sidereal time (LST), then it is expected the galactic emission to have the peak amplitude. This feature can be seen in both the data and simulation. These results confirm the first detection of the Galactic background by the GRAND Collaboration at high frequencies using the G@A prototype.

\section{G@A's First Candidate Cosmic Ray Event}
\label{sec:event}

\begin{figure}
    \centering
    \includegraphics[width=0.51\linewidth, trim={0 1.8cm 1.8cm 2.8cm},clip]{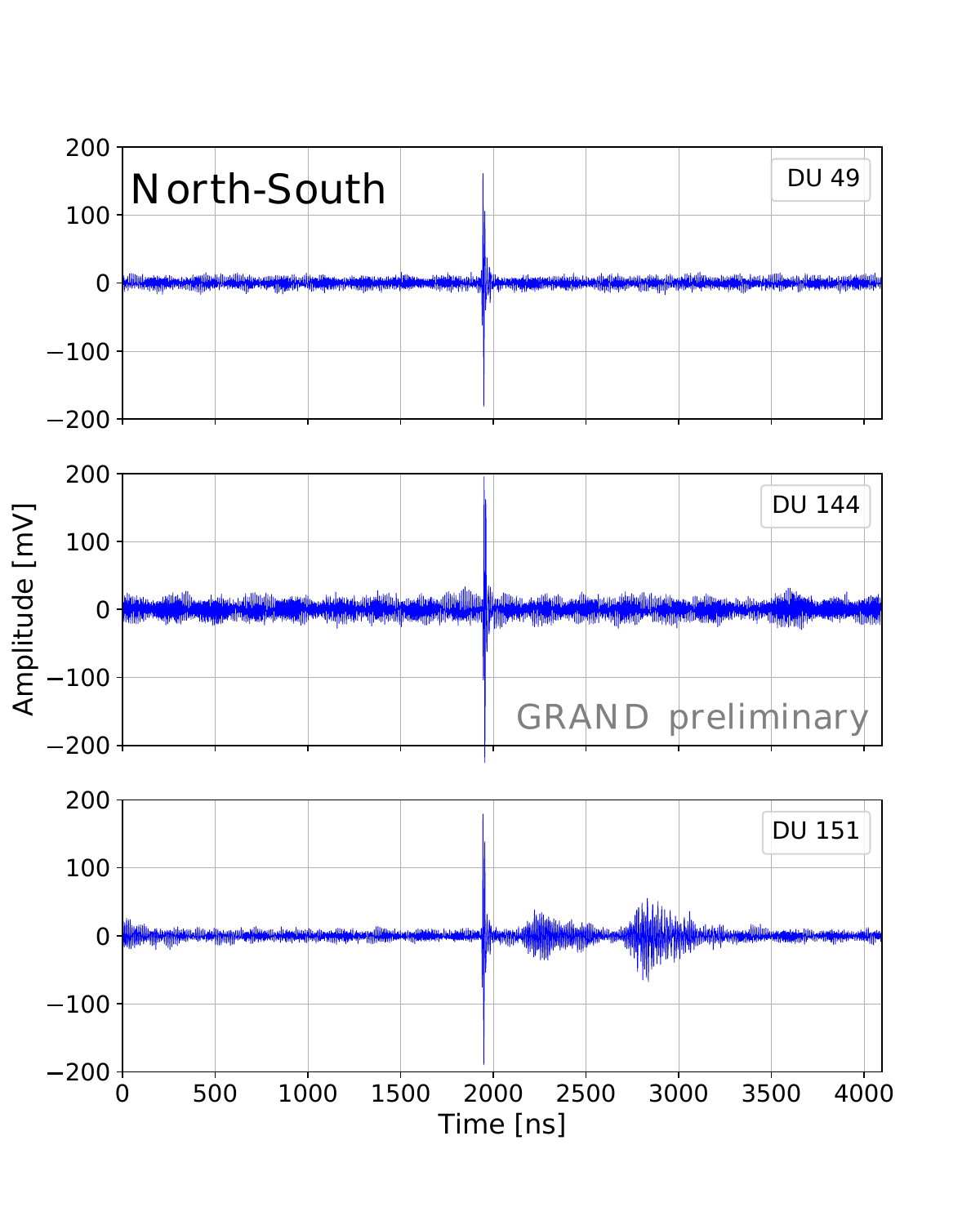}
    \includegraphics[width=0.483\linewidth, trim={1.0cm 1.8cm 1.8cm 2.8cm},clip]{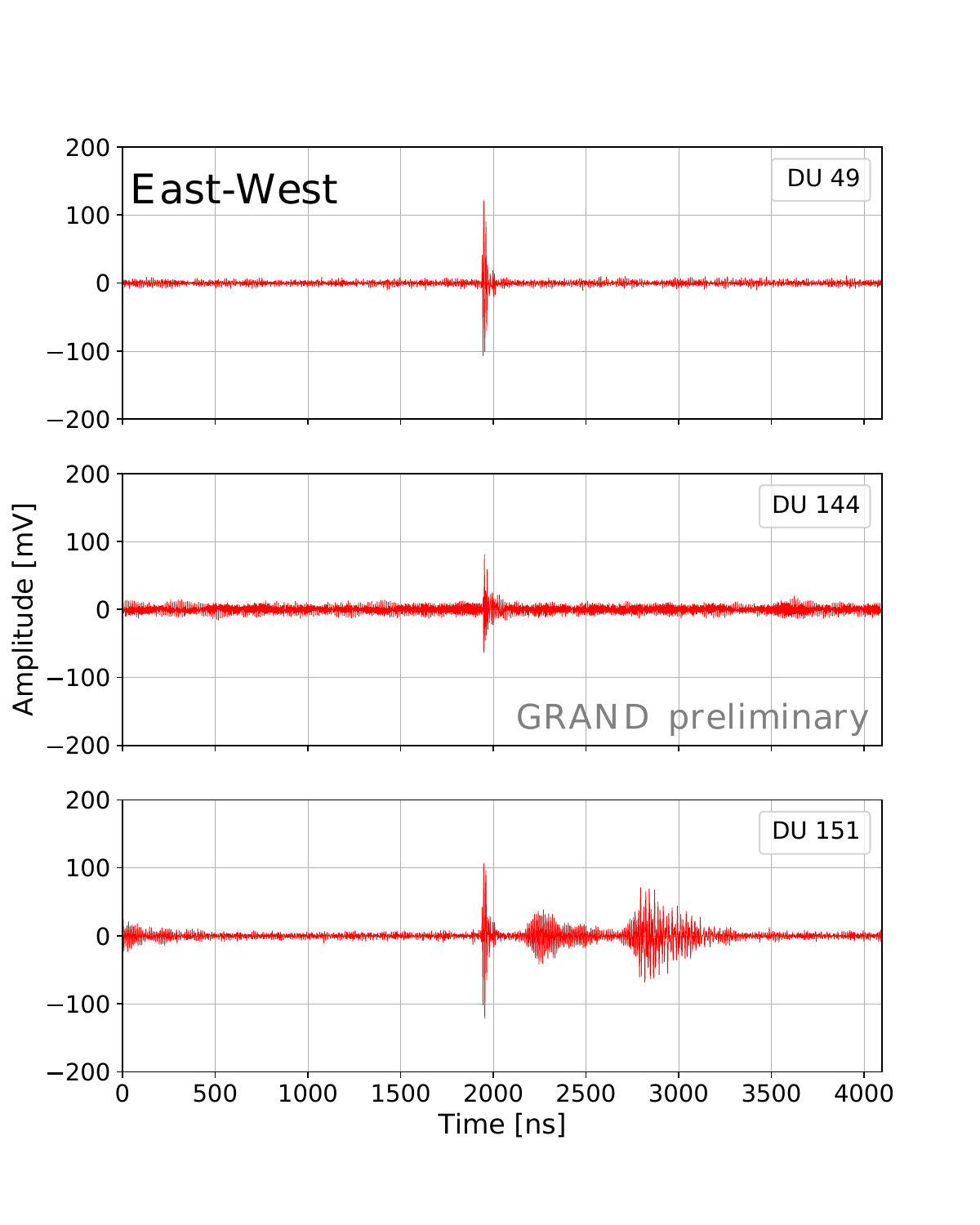}
    \caption{A self-triggered online coincidence was observed between three G@A detection units: 49, 144, and 151. The left panel shows the NS  polarization for the three units, while the right panel displays the EW  polarizations. 
    A coincident signal is also present in the vertical polarization of unit 49; however, the vertical channels in units 144 and 151 were not operational.}
    
    \label{fig:traces}
\end{figure}

The G@A  prototype has identified its first candidate cosmic ray (CR) event in coincidence with a Pierre Auger Observatory  triggered CR event. The G@A event occurred on September 16, 2024, at 20:06:08~UTC. It was detected as a  self-triggered coincidence event across three detection units (DUs): 49, 144, and 151. The G@A timestamps  for the coincident event are shown in black in Figure~\ref{fig:array} and the raw  voltage time-traces are shown in Figure~\ref{fig:traces}

 The G@A event, with three units in coincidence,  had its arrival direction reconstructed using an analytic Plane Wave Fit ~\cite{Ferriere2024}. The reconstruction results are as follows:
\[
     \phi = 143.4^\circ,\quad \theta = 77.59^\circ \quad \textrm{(Auger's coordinates),}
\]
\noindent for $\phi$ the azimuth angle, measured with respect to the east, and $\theta$ the zenith angle.

The corresponding Auger event was a T5 Surface Detector (SD) event detected at the same timestamp, with 22 stations in coincidence.  The Auger SD arrival direction reconstruction using a spherical wave fit yielded:
\[
\phi = 142.91^\circ \pm 0.02^\circ , \quad  \theta = 80.36^\circ \pm 0.05^\circ.
\]
The energy estimate from the Auger SD lateral distribution function (LDF) fit was:
\[
E = 1.39 \times 10^{19} \ \text{eV}.
\]

\vspace{-10pt}
\begin{figure}[ht]
 \centering
 \includegraphics[width=0.65\textwidth]{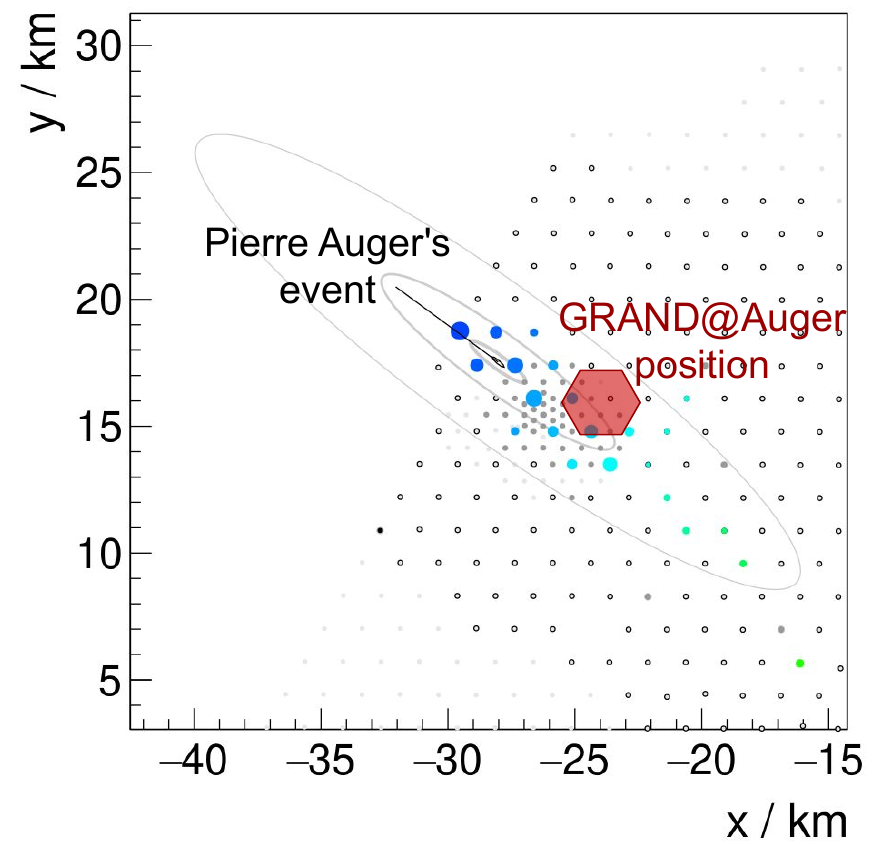}
 \caption{Coincidence event observed by the G@A prototype and the Pierre Auger Observatory on September 16, 2024, at 20:06:08~UTC. The G@A event, detected by three self-triggered GRAND units, shows a reconstructed arrival direction of $\theta = 77.6^\circ$, $\phi = 143.4^\circ$ (plane wave fit). The Auger event, detected by 22 surface stations, has a reconstructed arrival direction of $\theta = 80.4^\circ \pm 0.05^\circ$, $\phi = 142.9^\circ \pm 0.02^\circ$ (spherical wave fit) and an estimated energy of $1.39 \times 10^{19}$~eV. The close match in time and direction strongly supports a common cosmic ray origin. }  \label{fig:event}
 \end{figure}

A time offset of approximately 10 microseconds was observed between the G@A and Auger detections, with the G@A event triggering slightly earlier. The spatial and temporal coincidence, as shown in Figure \ref{fig:event}, along with the directional agreement from the reconstructions, provides strong evidence that this event is a cosmic ray detection by G@A. This is  the first high-energy cosmic-ray event observed in coincidence by GRAND and Auger.  It is also an important milestone  for GRAND, which demonstrates the self-trigger capabilities of G@A and its alignment with the independent Auger reconstruction.

\section{Conclusions}

The G@A prototype has demonstrated stable data-taking.
Self-triggered radio arrays like G@A require a precise understanding of the local electromagnetic background to optimize trigger settings and distinguish air-shower-like signals from real air-shower events. The detection of  self-triggered events suggests that the trigger configuration is functioning as expected and reasonably well understood.

A detailed data cleaning procedure was implemented to identify and mitigate anthropogenic noise sources such as television signals, beacon lines, and other localized radio interference. This effort has allowed the extraction of high-quality data for further analysis.

Regarding the Galactic background, while the data set was not long enough to clearly distinguish a daily period in GMT from the expected LST periodicity without simulation support, comparisons between the measured data and the simulations produced with \texttt{LFmap} show a good match, particularly at higher frequencies. In the 100--200~MHz band, the data suggest that the gain of the detector is understood to within approximately 10\% for the horizontal polarizations and about 50\% for the vertical channel. The G@A prototype has thus enabled the first detection of the Galactic background by GRAND at high frequencies and provided a unique measurement of the Galactic background at the Auger site in the 100--200~MHz frequency range.

Finally, the first candidate cosmic-ray event detected by G@A in coincidence with the Pierre Auger Observatory marks a significant milestone. Although the expected coincidence rate requires further investigation, the observed event, with an energy of $1.39 \times 10^{19}$~eV measured by Auger, shows strong indications of being a genuine cosmic-ray event. This is the first autonomous radio detection of a cosmic ray by GRAND. The detection demonstrates the potential of the GRAND technique and highlights the importance of cross-calibration with established observatories like Pierre Auger. These results pave the way for future studies and further advances in the development of large-scale, self-triggered radio arrays for the detection of ultra-high-energy particles.

{
}

\clearpage

\section*{Full Author List: GRAND Collaboration}

\scriptsize
\noindent
J.~Álvarez-Muñiz$^{1}$, R.~Alves Batista$^{2, 3}$, A.~Benoit-Lévy$^{4}$, T.~Bister$^{5, 6}$, M.~Bohacova$^{7}$, M.~Bustamante$^{8}$, W.~Carvalho$^{9}$, Y.~Chen$^{10, 11}$, L.~Cheng$^{12}$, S.~Chiche$^{13}$
% ...
\\
\\
$^{1}$Departamento de Física de Particulas \& Instituto Galego de Física de Altas Enerxías, Universidad de Santiago de Compostela, 15782 Santiago de Compostela, Spain \\
$^{2}$Institut d'Astrophysique de Paris, CNRS  UMR 7095, Sorbonne Université, 98 bis bd Arago 75014, Paris, France \\
$^{3}$Sorbonne Université, Université Paris Diderot, Sorbonne Paris Cité, CNRS, Laboratoire de Physique  Nucléaire et de Hautes Energies (LPNHE), 4 Place Jussieu, F-75252, Paris Cedex 5, France \\
$^{4}$Université Paris-Saclay, CEA, List,  F-91120 Palaiseau, France \\
$^{5}$Institute for Mathematics, Astrophysics and Particle Physics, Radboud Universiteit, Nijmegen, the Netherlands \\
$^{6}$Nikhef, National Institute for Subatomic Physics, Amsterdam, the Netherlands \\
$^{7}$Institute of Physics of the Czech Academy of Sciences, Na Slovance 1999/2, 182 00 Prague 8, Czechia \\
$^{8}$Niels Bohr International Academy, Niels Bohr Institute, University of Copenhagen, 2100 Copenhagen, Denmark \\
$^{9}$Faculty of Physics, University of Warsaw, Pasteura 5, 02-093 Warsaw, Poland \\
$^{10}$Key Laboratory of Dark Matter and Space Astronomy, Purple Mountain Observatory, Chinese Academy of Sciences, 210023 Nanjing, Jiangsu, China \\
$^{11}$School of Astronomy and Space Science, University of Science and Technology of China, 230026 Hefei Anhui, China \\
$^{12}$National Astronomical Observatories, Chinese Academy of Sciences, Beijing 100101, China \\
$^{13}$Inter-University Institute For High Energies (IIHE), Université libre de Bruxelles (ULB), Boulevard du Triomphe 2, 1050 Brussels, Belgium \\
% ...


\subsection*{Acknowledgments}

\noindent
The GRAND Collaboration is grateful to the local government of Dunhuag during site survey and deployment approval, to Tang Yu for his help on-site at the GRANDProto300 site, and to the Pierre Auger Collaboration, in particular, to the staff in Malarg\"ue, for the warm welcome and continuing support.
The GRAND Collaboration acknowledges the support from the following funding agencies and grants.
\textbf{Brazil}: Conselho Nacional de Desenvolvimento Cienti\'ifico e Tecnol\'ogico (CNPq); Funda\c{c}ão de Amparo \`a Pesquisa do Estado de Rio de Janeiro (FAPERJ); Coordena\c{c}ão Aperfei\c{c}oamento de Pessoal de N\'ivel Superior (CAPES).
\textbf{China}: National Natural Science Foundation (grant no.~12273114); NAOC, National SKA Program of China (grant no.~2020SKA0110200); Project for Young Scientists in Basic Research of Chinese Academy of Sciences (no.~YSBR-061); Program for Innovative Talents and Entrepreneurs in Jiangsu, and High-end Foreign Expert Introduction Program in China (no.~G2023061006L); China Scholarship Council (no.~202306010363); and special funding from Purple Mountain Observatory.
\textbf{Denmark}: Villum Fonden (project no.~29388).
\textbf{France}: ``Emergences'' Programme of Sorbonne Universit\'e; France-China Particle Physics Laboratory; Programme National des Hautes Energies of INSU; for IAP---Agence Nationale de la Recherche (``APACHE'' ANR-16-CE31-0001, ``NUTRIG'' ANR-21-CE31-0025, ANR-23-CPJ1-0103-01), CNRS Programme IEA Argentine (``ASTRONU'', 303475), CNRS Programme Blanc MITI (``GRAND'' 2023.1 268448), CNRS Programme AMORCE (``GRAND'' 258540); Fulbright-France Programme; IAP+LPNHE---Programme National des Hautes Energies of CNRS/INSU with INP and IN2P3, co-funded by CEA and CNES; IAP+LPNHE+KIT---NuTRIG project, Agence Nationale de la Recherche (ANR-21-CE31-0025); IAP+VUB: PHC TOURNESOL programme 48705Z. 
\textbf{Germany}: NuTRIG project, Deutsche Forschungsgemeinschaft (DFG, Projektnummer 490843803); Helmholtz—OCPC Postdoc-Program.
\textbf{Poland}: Polish National Agency for Academic Exchange within Polish Returns Program no.~PPN/PPO/2020/1/00024/U/00001,174; National Science Centre Poland for NCN OPUS grant no.~2022/45/B/ST2/0288.
\textbf{USA}: U.S. National Science Foundation under Grant No.~2418730.
Computer simulations were performed using computing resources at the CCIN2P3 Computing Centre (Lyon/Villeurbanne, France), partnership between CNRS/IN2P3 and CEA/DSM/Irfu, and computing resources supported by the Chinese Academy of Sciences.

\end{document}